\documentstyle[prl,aps]{revtex}
\begin{document}
\draft
\twocolumn[\hsize\textwidth\columnwidth\hsize\csname @twocolumnfalse\endcsname

\title{Microscopic Derivation of Magnetic Flux Density Profiles, Magnetization
Hysteresis Loops, and Critical Currents in Strongly Pinned Superconductors}

\author{C.~Reichhardt, C.~J.~Olson, J.~Groth, Stuart Field, and Franco Nori}
\address{Department of Physics, The University of Michigan, Ann Arbor,
Michigan 48109-1120}

\date{\today}
\maketitle
\begin{abstract}
We present a microscopic derivation, without electrodynamical assumptions,
of $B(x,y,H(t))$, $M(H(t))$, and $J_c(H(t))$, in agreement with experiments
on strongly pinned superconductors, for a range of values of the density
and strength of the pinning sites.
We numerically solve the overdamped equations of motion 
of these flux-gradient-driven
vortices which can be temporarily trapped at pinning centers.
The field is increased (decreased) by the addition (removal) of flux lines
at the sample boundary, and
complete hysteresis loops can be
achieved by using flux lines with opposite orientation.
The pinning force per unit volume we obtain for strongly-pinned vortices,
$J_c B \sim n_p f_p^{1.6}$, interpolates between the following two extreme
situations: very strongly-pinned independent vortices, where
$J_c B \sim n_p f_p$, and the 2D Larkin-Ovchinikov collective-pinning
theory for weakly-pinned straight vortices, where $J_c B \sim n_p f_p^{2}$.
Here, $n_p$ and $f_p$ are the density and maximum force of the pinning sites.
\end{abstract}
\vskip2pc]
\narrowtext  

\section{Introduction}

Flux distributions in type-II superconductors are commonly inferred from
magnetization and critical current measurements\cite{reviews}
and interpreted in the context of the Bean model\cite{bean} or its variations.
The Bean model, which has been widely used for over three decades,
{\it postulates\/} that the current density in a hard superconductor
({\em i.e.}, with strong pinning) can only have three values:
$-J_c$, $0$, and $+J_c$, where $J_c$ is the critical current density,
which is independent of the local magnetic flux density ${\bf B}(x,y,t)$.
The Bean model and its many variants 
make no specific claims with regard to the {\it microscopic\/} mechanism
controlling the trapping of vortices.  Bean's postulate, $J_c=$constant,
was modified several times by Kim {\em et al.}\cite{kim}:
$J_c \sim 1/ B \ $ [3$^a$];              
$J_c \sim 1/(b_0+B) \ $ [3$^b$,3$^c$];  
$J_c \sim 1/(b_0+B+b_2B^2+b_3B^3+\ldots) \ $  [3$^b$]; 
where $b_i$ are constants.
On the other hand, Fietz {\em et al.} \cite{fietz64} suggested that
$J_c \sim \exp(-B/b_0) \ $;
while Yasuk\={o}chi {\em et al.} \cite{yas} suggested
$J_c \sim 1/B^{1/2}$.
These, and other proposals made during the 1960s, were followed by several
other phenomenological modifications of $J_c(H)$ during the following two
decades\cite{reviews,zeldov}.
A microscopic description,
{\it without\/} assuming any particular $B$-dependence of $J_c$,
of these flux distributions---in terms of interacting vortices and
pinning sites---can be very valuable for a better understanding of
commonly measured bulk quantities.

One of the most effective methods of investigating the microscopic behaviour
of flux in a hard superconductor is with computer simulations
(see, e.g., \cite{simulations,ray94}, and references therein).
In this paper, we present molecular dynamics (MD) simulations of the evolution
of rigid flux lines in a hard superconductor. We first introduce our model for
vortex-vortex and vortex-pin interactions as well as the corresponding
antivortex interactions. We then investigate the flux profile which results
from a varying applied field; from such flux profiles we obtain full
hysteresis loops indicating that our model has the essential microscopic
ingredients underlying the experimentally measured macroscopic quantities.
We also investigate the behaviour of $J_c(H)$ for a controlled
range of pinning parameters.

\section{Simulation}

Our simulation geometry is that of an infinite slab of superconductor in
a magnetic field applied {\em parallel} to the slab surface.  Thus,
demagnetization effects are unimportant.  We also treat the vortices as
perfectly stiff, so that we need to model only a two-dimensional (2D)
slice of the 3D slab.  Our system is periodic in the plane
perpendicular to the applied field, and we measure distances in units of
the penetration length $\lambda$.  Here, we present results for a system
of size $36\lambda \times 36\lambda$.  The simulation, described in
further detail below, consists of slowly ramping an external magnetic
field.  Flux lines enter the edge of the sample and their positions are
allowed to evolve according to a $T=0$ MD algorithm.  The resulting vortex
distributions at any external field can then be deduced as a function
of distance into the sample.

\subsection{Sample Geometry and Time-Dependent Field}

The actual sample region is heavily pinned, and extends from position
$x=6\lambda$ to $x=30 \lambda$ (Fig.~1).  Outside the sample itself is a
region with no pinning which extends from
$x = 0  \lambda $ to $x=6  \lambda $ and from
$x = 30 \lambda $ to $x=36 \lambda $
(with $ 36\lambda = 0\lambda $ according to our periodic boundary conditions).
\ This sample geometry is shown in the upper panels of Fig.~1.
Here, the sample (pinned) region occupies the central $2/3$ of the system,
and the unpinned region the outer $1/3$.

We simulate the ramping of an external field by the slow addition of
flux lines to the outside unpinned region.  Because there is no pinning
in this region, the flux lines there will attain a fairly uniform
density, and we may define the applied field $H$ as $\Phi_0$ times this
density.  Flux lines from the external region will move into the sample
through points at the sample edge where the local energy---as
determined by the local pinning and vortex interaction---is low.
Thus, our simulation 
models the real situation where vortices nucleate at such low-energy
regions at the surface.

Further, in a real superconductor, vortices near the surface are not
expelled by their interior neighbors because of a field-induced Meissner
current flowing at the surface.  Again, our external ``bath'' of
vortices simulates this behavior by providing a balancing inward force,
proportional to the external field, on those vortices near the sample
boundary.

\subsection{Equations of Motion}

The force per unit length \cite{reviews} between two vortices located at
${\bf r}_i$ and ${\bf r}_j$   is
\begin{equation}
\ f^{vv}  = \frac{ \Phi_0^2 }{ 8 \pi^2 \lambda^3 }
\ K_1 \left(   \frac{ | {\bf r}_i - {\bf r}_j | }{ \lambda } \right) \, . \
\end{equation}
We model the vortex-vortex force interaction in its exact form by using the
modified Bessel function $K_1$.  This force decreases exponentially at
distances larger than $\lambda$, and we cut off the (by then negligible)
force at distances greater than $6\lambda$.  Further, we have cut off the
logarithmic divergence of the force for distances less than $0.1\lambda $.
These cutoffs were found to produce negligible effects on the dynamics for
the range of parameters investigated. Thus, the force (per unit length)
on vortex $i$ due to other vortices (ignoring cutoffs) is
$
\ {\bf f}_{i}^{vv} = 
\ \sum_{j=1}^{N_{v}} \ f_{v} \ K_{1}(
|{\bf r}_{i} - {\bf r}_{j}| / \lambda  ) \
{\bf {\hat r}}_{ij} \, .
$
Here, the ${\bf r}_{j} $ are the positions of the $N_v$ vortices within a
radius $6\lambda$,
$ \ {\bf {\hat r}}_{ij} =
({\bf r}_{i} - {\bf r}_{j}) / |{\bf r}_{i} - {\bf r}_{j}| $,
$\ f_v = \pm f_0$, and
\begin{equation}
\ f_0 = \frac{ \Phi_0^2 } {8 \pi^2 \lambda^3} \ .
\end{equation}
The sign of the interaction is determined by $ f_{v} $;
we take
$ f_{v} = +f_0 $ for repulsive vortex-vortex interactions and
$ f_{v} = -f_0 $ for attractive vortex-antivortex interactions.  A vortex
and antivortex annihilate and are removed from the system if they come
within $0.3 \lambda $ of one another
\cite{reviews}.
Forces are measured in units of $f_0$, lengths in units of $\lambda$, and
fields in units of $\Phi_0/\lambda^2$.

We model the pinning potential\cite{pinning} as $N_p$ short-range parabolic
wells at positions ${\bf r}_k^{(p)}$.
The equation of motion for a vortex moving with velocity $v$ is
$f=\eta v$, where $\eta$ is the viscosity
($\approx \Phi_0 H_{c2}/\rho_n$, with $\rho_n$ being the normal-state
resistivity).
Thus, the overall equation for the overdamped motion of a vortex subject to
vortex-vortex and pinning forces is
\begin{equation}
\ {\bf f}_{i} = {\bf f}_{i}^{vv} + {\bf f}_{i}^{vp} = \eta {\bf v}_{i} \ ,
\end{equation}
where
\begin{eqnarray}
  {\bf f}_{i} & = & 
\ \sum_{j=1}^{N_{v}}\, f_{v} \,
K_{1} \left( \frac{ |{\bf r}_{i} - {\bf r}_{j}| }{ \lambda } \right)
\, {\bf {\hat r}}_{ij}
\nonumber \\  & + & \sum_{k=1}^{N_{p}} \frac{f_{p}}{\xi_{p}}
\ |{\bf r}_{i} -  {\bf r}_{k}^{(p)}| \
\Theta\left( \frac{ \xi_{p} - | {\bf r}_{i} - {\bf r}_{k}^{(p)} | }{\lambda}
\right) \ {\bf {\hat r}}_{ik} \, .
\end{eqnarray}
Here, $\Theta $ is the Heaviside step function,
$ \xi_{p} $ is the range of the pinning potential, and $f_{p}$ is the strength
(maximum pinning force) of each well, measured in units of $f_0$.
For all the simulations presented here $\xi_{p} = 0.12\lambda$ and $\eta=1$.
The parameters we vary here are the pinning strength $f_p$
and the average distance between pinning sites $d_p$
(which determines the pinning density $n_p$ via $n_p=1/d_p^2$).
Many other parameters can be varied, making the systematic study of this
problem very complex.  A more thorough investigation with
different pinning-potential ranges, pinning potential-shapes,
non-uniform strength distributions, and non-random pinning positions
will be presented elsewhere.
Here, the pinning sites have uniform strengths and are placed in the sample
at random, but non-overlapping, positions.  The pinning strength $f_{p}$
is varied from $0.2 f_0$ to $1.0 f_0$, and $d_p$ is varied
from $\lambda/3$ to $\lambda\ $ (i.e., the pin density $n_p$ varies
from $1/\lambda^{2}$ to $9/\lambda^{2}$).

\section{Magnetic Flux Density Profiles}
Several general features of our simulations are shown in Fig.~1.  In
the upper frame of Fig.~1a, we show a top view of the vortex positions
after the external field has been ramped up from zero.  As we have
stated, this external field is represented by the vortices in the
unpinned regions to the left and right of the central, pinned, sample
region.  Here, vortices have been added to the unpinned region to a
final density of about $1.2$ vortices/$\lambda^2$; since each vortex
carries a flux $\Phi_0$, this corresponds to a magnetic field of
$1.2 \ \Phi_0/\lambda^2$.  For a real superconductor with a penetration depth
of, e.g., $1000$\AA , this corresponds to $H=2.5$ kOe.

We note in Fig.~1(a) that many of the vortices added to the unpinned region
have been forced into the central sample region at this stage.  They do
not do so uniformly due to the presence of 3456 pinning sites (not shown),
with a typical intersite distance of $\lambda/2$ and $f_p=0.9 f_0$.
We see the characteristic
density gradient determined by a balancing of the vortex-vortex forces
with the local pinning forces.  Since this gradient was achieved in our
simulation solely by the slow ramping of an external magnetic field, we
have obtained the field profiles inside a pinned superconductor using
only {\em microscopic} information such as vortex-vortex and vortex-pin
interactions.  We should also contrast our simulations with those
modeling {\em current-driven\/} vortices.  In such simulations the driving
force on each vortex is somewhat artificially modeled by an externally-imposed
``uniform'' current.
Our simulation correctly models the driving force as a result of local
interactions.

The lower frame of Fig.~1a shows the resulting flux density profiles,
found by averaging the vortex density over slices parallel to the sample
edges.
Such profiles clearly show the essentially constant flux density
in the external regions, and the detailed nature of the flux gradient
within the sample.
Of course, these profiles may be obtained at any value of the external field.
Figure~1b shows the system after the external field
has been ramped {\em down} from a high value to zero.
The small field outside the sample is an artifact due to the smearing
of the vortex fields.
Now, flux remains trapped within the sample and the field gradient has
changed sign.
We notice that near the sample edges, where the field is small,
the gradient in the flux density is quite large.  Thus our simulation
correctly models the increase in flux gradient (or, equivalently,
critical current) at low fields, where intervortex interactions are weak
and pinning dominates.

In Fig.~2 we show flux density profiles for a complete cycle of
the field, with the same sample parameters as in Fig.~1.  During the initial
ramp-up stage (Fig.~2, left), we increase the external field from
zero to a final value of about $1.9\  \Phi_0/\lambda^2$.  We see the evolution
of the internal flux profile from first penetration at low fields, to
the first complete penetration at a field $H^* \approx 0.8\ \Phi_0/\lambda^2$,
to higher values of $B$ at larger $H$.  We again note the flux gradient is
quite high at low fields, but becomes flatter---and less
field-dependent---at high fields.


%

Of course, in real superconductors no vortices will enter the sample until
$H > H_{c1} \approx (\ln \kappa / 4 \pi ) (\Phi_0/\lambda^2) $,
where $\kappa=\lambda/\xi$.
However, for $\kappa$'s in the wide physically relevant range from $2$ to
$100$, $H_{c1}$ varies from
$0.05 \; \Phi_0/\lambda^2$ to $0.36 \;\Phi_0/\lambda^2$.
Thus, $H_{c1}$ is small in the range of fields we explore.
In any event, since we are only interested in the {\it mixed\/} state
and not the Meissner phase,
we will work in the approximation where $H_{c1}$ is negligible.
%


During the ramp-down stage
(Fig.~2, center), the field is lowered through zero to large
{\em negative} values.  The ramping down is initially effected
by simply removing vortices from the unpinned
region.  However, after the external field reaches zero, it is reversed
by the addition of {\em antivortices} in the unpinned region.  During
the beginning of this ramp-down stage, we note the appearance of the
characteristic ``gull-wing'' flux profile as the internal remnant flux
located close to the sample edges begins to be removed.
Notice that at external fields near zero
the internal field hardly changes at all as the external field is swept.
This is again because of the very steep gradients possible near zero field,
where pinning dominates.  Thus, the effect of a change in an external field
near zero propagates only a very small distance into the sample.

As the field decreases below $H=0$ (in Fig.~2, center),
$B(x)$ continues to have its $\wedge$-shaped 
profile.  We note that for small negative fields the sample contains both
vortices {\em and\/} antivortices.  However, the pinning for both types is
attractive, and so they remain locally trapped and annihilate only when
their mutual attraction overcomes the pinning.  This only occurs when they
are closely spaced, within $0.3 \lambda$. Finally, in the last ramp-up stage
(Fig.~2, right), the full cycle is completed by increasing the field
from the large negative value up to a large positive field, where the flux
profile looks identical to the initial ramp-up stage of the cycle.

One clear advantage of our simulation is that we can obtain direct
{\it spatio-temporal\/} information on the distribution of flux {\em inside}
the sample. However, experimentally this is quite difficult, especially for
bulk samples.  Instead, average quantities, like magnetization curves, are
typically obtained.  From the field cycles shown in Fig.~2, we can easily
obtain such magnetization loops from our simulation.  Further, in our
simulation it is simple to vary microscopic parameters such as pin density and
strength.  Thus, our simulations allow for a systematic study of the
dependence of {\it macroscopic measurements}, such as the magnetization, on
{\it microscopic system parameters}.  It may also be possible to use our
results in the reverse problem, so that some understanding of the microscopics
of the pinning \cite{pinning} may be obtained from experimentally
determined macroscopic measurements.

\section{Magnetization Hysteresis Loops}

Experimentally, what is typically measured is the average magnetization
over the sample volume.  In our simulation, we thus calculate the
average magnetization
\begin{equation}
\overline{M} = \frac{1}{4 \pi V } \int (H - B) \ dV \ .
\end{equation}
In Fig.~3 we construct magnetization loops as two key sample microscopic
parameters---the pinning density and strength---are varied.  Fig.~3a
shows complete magnetization loops obtained with the density of pins
held constant at $4/\lambda^2$, but at three different values of the
pinning strength $f_{p}$.
One can see clearly that by increasing the pinning strength the hysteresis
loops become much wider.  This is because a large pinning force yields a
large field gradient.  Thus $\overline{M}$, which is essentially the difference
between the internal and external fields, will be larger for large
$f_{p}$.  For instance, the remnant $\overline{M}$ is larger for stronger
pinning.
The $\overline{M}(H)$ loops all show a maximum when the external field is small
($H \leq H^*$) and close to $H^*$. This again is due to the pinning being
most effective for low fields ($H \leq H^*$).
Figure 3b shows magnetization loops obtained for several pinning
densities. Experimentally, one may systematically vary this parameter by
the introduction of columnar defects using irradiation \cite{reviews,civale}.

\section{Critical Current versus pinning density and strength}

Although magnetization loops are very useful for comparison with
experimental data, we have emphasized that our simulations allow us to
directly compute the local flux distribution inside the sample.  Thus,
we may directly measure the local critical current density $J_c$ using
Maxwell's equation $dB/dx = \mu_0 J$.  At every point on flux density
profiles such as Fig.~2 we may compute the local slope ($= dB/dx$) and
the corresponding local field $B$.  This allows us to determine a large
number of values of $J_c(B)$.  We then bin these values to obtain
suitably averaged curves of $J_c$ vs. $B$.

As we have discussed, there
are in the literature a great variety of functional dependences of $J_c$
on $B$, corresponding to different {\em ad hoc} electrodynamical
assumptions.  The original Bean model predicts $J_c$ to be independent
of $B$.  The varying slopes of the flux density in Fig.~2 show that this
prediction is not borne out in our simulation (except at relatively high-fields
where the vortex-vortex force dominates; e.g.,
for weak-pinning samples with $\lambda^2 n_p = 4.0$, $f_p=0.2f_0$).
Kim {\em et al.}\cite{kim} have proposed that the critical current depends
on $B$ as
\begin{equation}
\alpha = J_c (B + b_0) \ ,
\end{equation}
where $\alpha$ is field-independent and has units of force per unit volume.
In this model, plots of $1/J_c$ vs. $B$ should appear as straight lines
with slopes $1/\alpha$ and intercept $b_0/\alpha$.  The physical
interpretation of the constant $b_0$ in Kim's model is unclear\cite{kim}.

In Fig.~4 we plot $1/J_c$ vs. $B$, with $J_c$ determined from our flux
density plots during the initial ramp up phase.  We plot $1/J_c$ for
several realizations of the pinning density $n_p$ and strength $f_p$.
Fig.~4a shows $1/J_c$ vs $B$ for four different field sweeps with the
pinning density varied from $1.0/\lambda^2$ to $9.0/\lambda^2$; in
Fig.~4b we vary the pinning strength from $0.2 f_0$ to $0.9 f_0$.  Over a
large region of the field, we find that $1/J_c$ is indeed linear in
field, as in Kim's model.  We can then fit the linear portions of each
curve to straight lines as shown, and extract the inverse slope
$\alpha$.  For fields such that $B \gg b_0$, Kim's relation reads
$\alpha \approx J_c B$ which is the Lorentz force per unit volume. Since
this force is exactly balanced by the pinning force, we can interpret
$\alpha$ as the maximum pinning force per unit volume.  $b_0$ is
typically in the range of $0.4$ to $0.7$ $\Phi_0/\lambda^2$, but even below
$b_0$, $\alpha$ is clearly a measure of the relative effectiveness of
the pinning.

In the inset to Fig.~4a, we plot the values of $\alpha$ determined from the
slopes of the $1/J_c$ curves as a function of the pinning strength $f_p$
or density $n_p$.  The pinning force per unit volume has an approximate
linear rise with $n_p$, and the curve with dark triangles follows
$\alpha \sim f_p^{1.6} $
(if we assume that $\alpha=0$ when $f_p=0$).
\ Even though the vortex dynamics in our samples is not dominated by
elastic flow and collective weak-pinning, it is interesting to compare
these results with the predictions of the Larkin-Ovchinnikov
(LO)\cite{LO} collective-pinning theory---where weakly-pinned
vortices interact elastically inside a typical correlated volume.
The 2D LO prediction for rigid vortices becomes
\begin{equation}
J_c B \sim n_p f_p^2 \ ,
\end{equation}
which is somewhat different from
\begin{equation}
J_c B \ \sim n_p f_p^{1.6} \ \, ,
\end{equation}
obtained from our strongly pinned vortices.
The opposite regime of the LO weakly-pinned collective vortices
is given by the very strongly-pinned independent vortices where
\begin{equation}
J_c B \ \sim n_p f_p^{1} \ \, .
\end{equation}
Thus, our results indicate that our vortices are in an intermediate
state between the two extreme regimes described above.

We plot our values for $J_c$ in practical SI units.  The weakest pinning
in our simulation occurs at our highest fields,
where $1/J_c$ is about $100 \mu_0 \lambda^3/\Phi_0$.
For a $\lambda$ of $1000$ \AA, this
corresponds to a critical current $J_c = 1.6 \times 10^6$ A/cm$^2$,
which is in practice a very reasonable value.
Our highest critical
currents, at low fields and high pin strength or density, are about a factor
of ten higher.  Thus, our parameters generally appear to model 
realistic materials.


\section{Conclusions}

To summarize, we have perfomed molecular-dynamics simulations of
vortices and antivortices interacting with a controlled range of
pinning strengths and densities.  In these simulations we have only
considered vortex-vortex and vortex-pin interactions;
{\it no\/} extra force was needed to simulate a Lorentz force.
Thus, our results show that the Lorentz force can be considered as a
consequence of a flux gradient arising strictly from the interactions of
vortices and pins.  We compute the flux density profile that develops with
a varying applied field, for both vortices and anti-vortices as the external
field is cycled through a loop.  Our computed
complete hysteresis loops show realistic behaviour
with varying pinning strength and density, indicating that our model
contains the essential physics. We have obtained $J_{c}(H) $ by focusing on
the flux gradient that develops naturally from the vortex-pin interactions
and find that it monotonically decreases with an increasing external field
with the fall off determined by the microscopic pinning parameters.

\section{Acknowledgements}

This work was supported in part by the NSF under grant No.~DMR-92-22541,
and by SUN microsystems.

\newpage

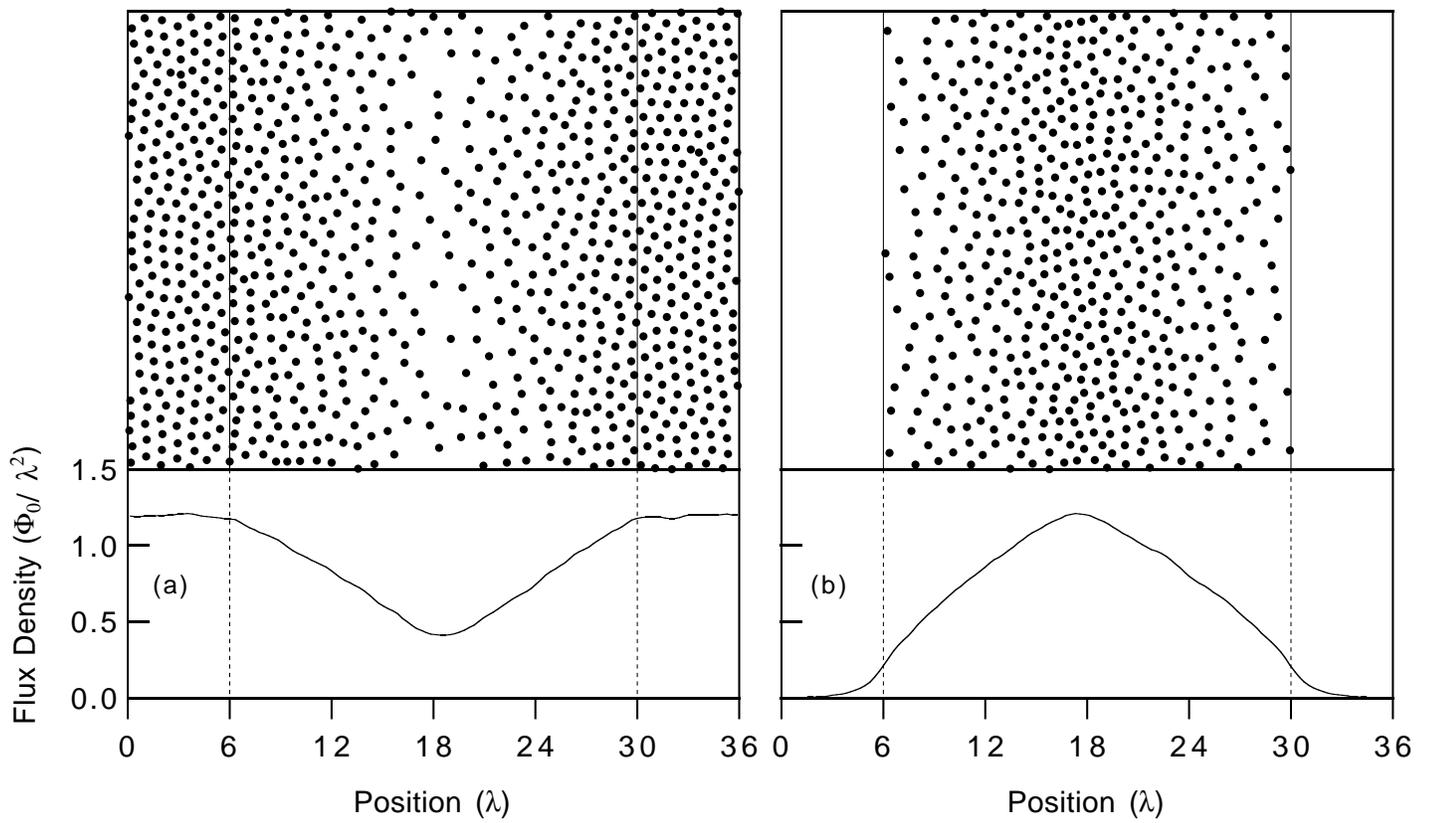
\begin{figure}
\caption{Top view of the region where flux lines, indicated by dots, move.
(a) Snapshot during the initial ramp-up phase,
(b) snapshot of the remnant magnetization after ramping down the external
field. The bottom panels show
$B(x)= (36 \lambda)^{-1} \int_{0}^{36\lambda} d\!y \, B(x,y)$,
i.e., the flux density profile versus $x$, averaged over the vertical
direction $y$.
The $24 \lambda \times 36 \lambda$ sample has $3456$ pinning sites, and
$f_p=0.9f_0$.
} \label{fig1}\end{figure}

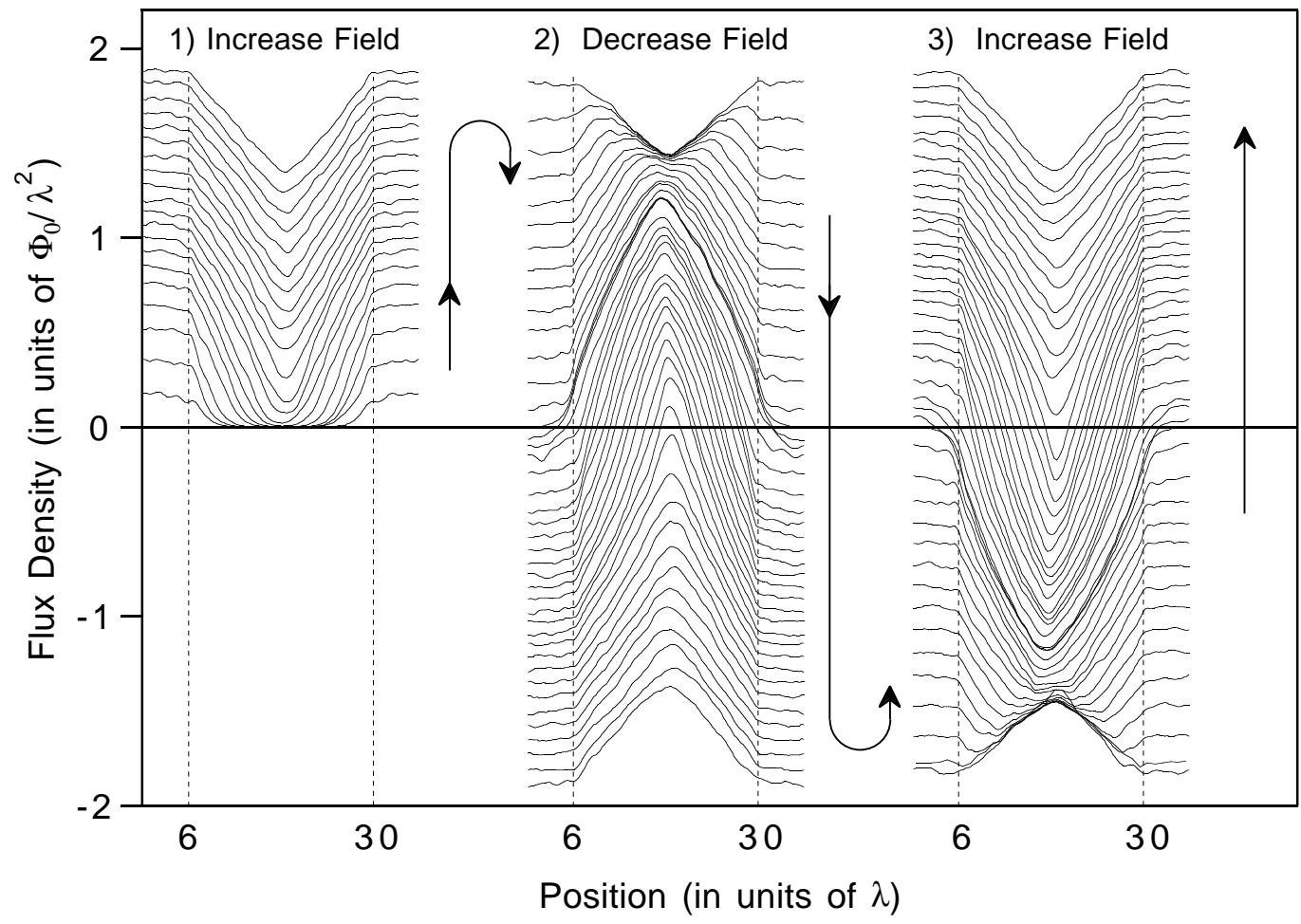
\begin{figure}
\caption{Magnetic flux density profiles $B(x)$ for the
(1) initial ramp-up phase,
(2) ramp-down stage reaching a negative field, and
(3) final ramp-up phase, for the same sample described in Fig.~1 and the text.
The flat plateaus on either side of the sample show the density in the
unpinned region, mimicking the external field, and the jagged
$\vee$-- and $\wedge$--shaped profiles correspond to the flux density in the
pinned region.}
\label{fig2}\end{figure}

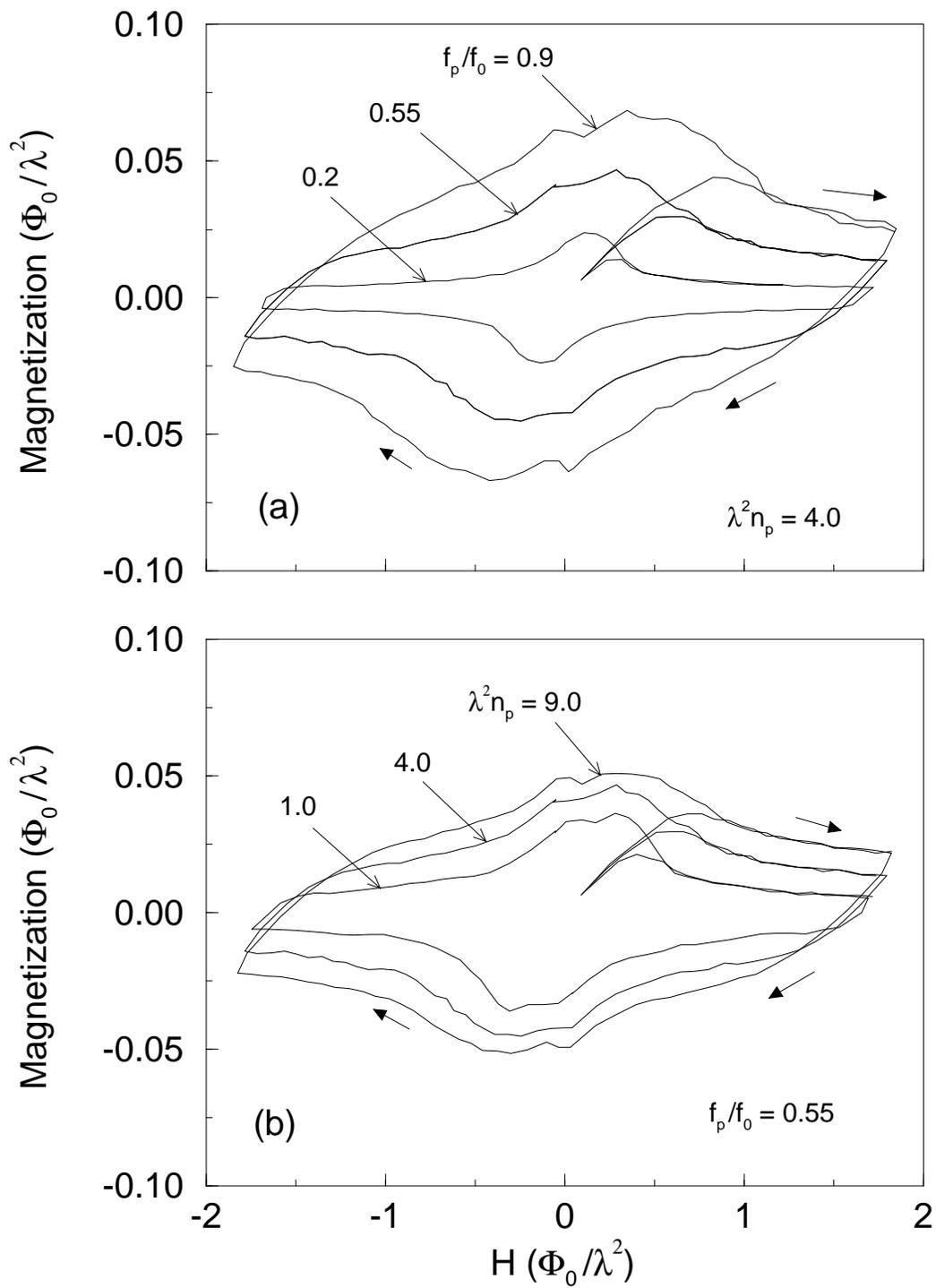
\begin{figure}
\caption{Magnetization hysteresis curves $M(H(t))$.  In (a) the maximum
pinning force is varied ($f_p=0.9f_0$, $0.55f_0$, $0.2f_0$) for a fixed
average distance between randomly distributed pinning sites, $d_p=\lambda/2$
(i.e., $\lambda^2 n_p=4$).
\ In (b) the pinning-site density $n_p$ is varied while $f_p=0.55 f_0$.
A higher value of $f_p$ and/or $n_p$ increase $J_c$
($\sim$ width of the $M(H)$ hysteresis loop)
in the manner shown in Fig.~4.
For each $M(H)$ loop shown, the maximum number of flux lines inside the
pinned sample is about 1000.
}
\label{fig3}\end{figure}

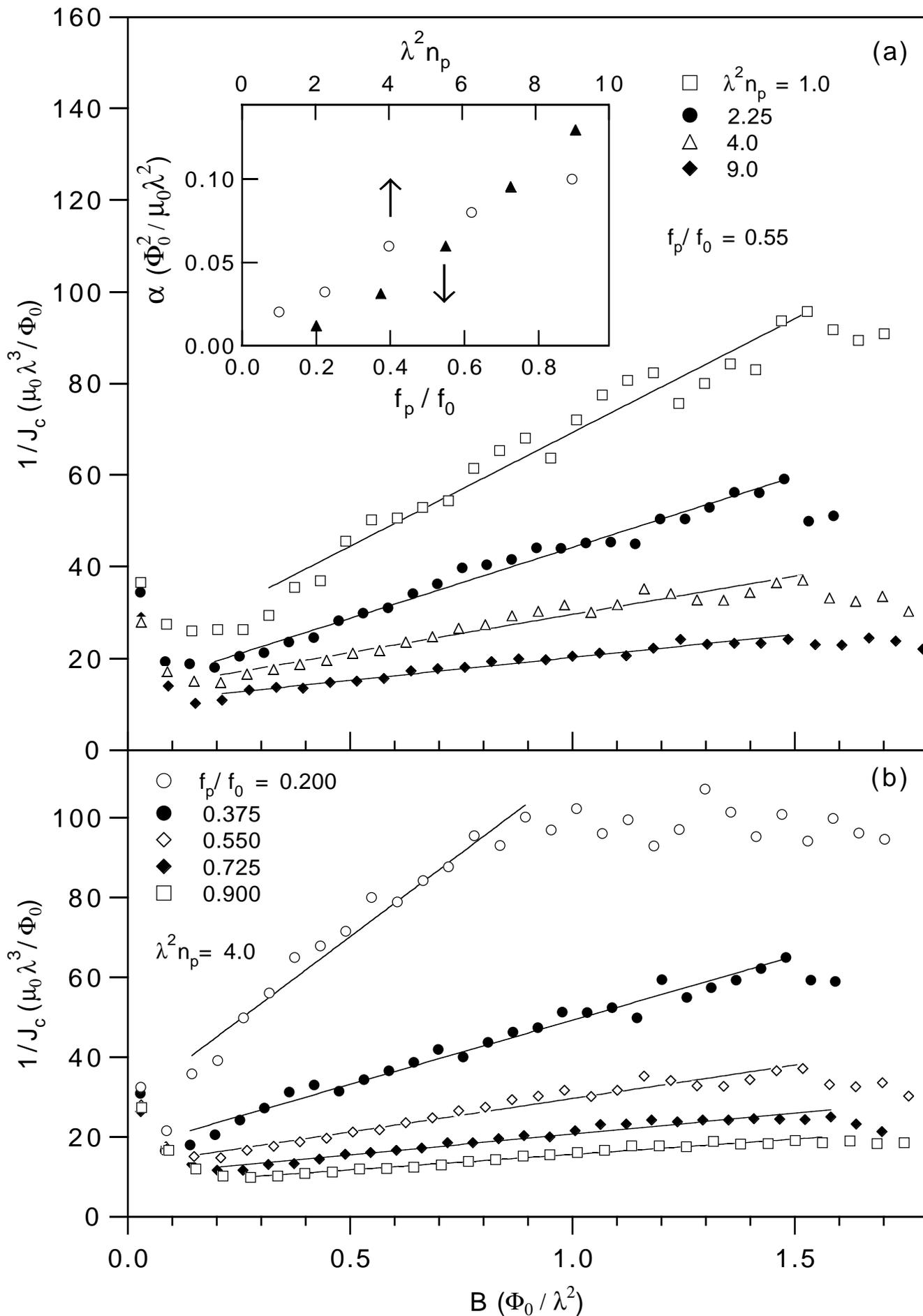
\begin{figure}
\caption{(a) $1/J_{\rm c}(B) $ for several values of the pin density $n_p$
(and fixed pinning-site strength $f_p$);
(b) $ 1/J_{\rm c}(B) $ for several values of $f_p$
(and fixed $n_p$).
The insets show the dependence of the maximum pinning force $\alpha$ on
$f_p$ (dark triangles) and on $n_p$ (open circles)
The values of $\alpha$ are obtained from the (solid line) linear fits
shown in the larger panels.}
\label{fig4}\end{figure}

\end{document}